\documentclass[preprint,prb,bibnotes,superscriptaddress,showkeys]{revtex4-1}
\usepackage{hyperref}
\usepackage{graphicx}
\usepackage{amssymb, amstext, amsmath}
\usepackage{color}
\usepackage{verbatim}
\usepackage[version=3]{mhchem} % Formula subscripts using \ce{}
\begin{document}

\title{Can disorder enhance incoherent exciton diffusion?}
\author{Elizabeth M. Y. Lee}
\author{William A. Tisdale}
\affiliation{Department of Chemical Engineering, Massachusetts Institute of Technology, Cambridge, MA 02139}
\author{Adam P. Willard}
%\phone{+1 (617) 253-1480}
\email{awillard@mit.edu}
\affiliation{Department of Chemistry, Massachusetts Institute of Technology, Cambridge, MA 02139}

%\textbf{Graphical TOC Entry}
%\begin{figure}
%\includegraphics{TOC.pdf}
%\end{figure}


% 150 words max
%%%%%%%%%%%%%%%%%%%%%%%%%%%%%%%%%%%%%%%%%%%%%%%%%%%%%%%%%%%%%%%%%%%%%
\begin{abstract}

Recent experiments aimed at probing the dynamics of excitons have revealed that semiconducting films composed of disordered molecular subunits, unlike expectations for their perfectly ordered counterparts, can exhibit a time-dependent diffusivity in which the effective early time diffusion constant is larger than that of the steady state. This observation has led to speculation about what role, if any, microscopic disorder may play in enhancing exciton transport properties.  In this article, we present the results of a model study aimed at addressing this point. Specifically, we present a general model, based upon F\"orster theory, for incoherent exciton diffusion in a material composed of independent molecular subunits with static energetic disorder. Energetic disorder leads to heterogeneity in molecule-to-molecule transition rates which we demonstrate has two important consequences related to exciton transport. First, the distribution of local site-specific diffusivity is broadened in a manner that results in a decrease in average exciton diffusivity relative to that in a perfectly ordered film. Second, since excitons prefer to make transitions that are downhill in energy, the steady state distribution of exciton energies is biased towards low energy molecular subunits, those that exhibit reduced diffusivity relative to a perfectly ordered film. These effects combine to reduce the net diffusivity in a manner that is time dependent and grows more pronounced as disorder is increased. Notably, however, we demonstrate that the presence of energetic disorder can give rise to a population of molecular subunits with exciton transfer rates exceeding that of subunits in an energetically uniform material. Such enhancements may play an important role in processes that are sensitive to molecular-scale fluctuations in exciton density field. 

\end{abstract}


\maketitle
\section[intro]{Introduction}
Excitons are Coulombically bound electron-hole pairs that mediate the transport of energy in many molecular scale systems. The dynamics of excitons are fundamental to processes such as photosynthetic light harvesting~\cite{Engel2007,Olaya-Castro2012,Kassal2013,Chenu2015}, photocurrent generation in solar cells~\cite{Coakley2004,Mayer2007,Nozik2008,Servaites2011,Menke2012,Graetzel2012}, and photoluminescence in light emitting diodes (LEDs)~\cite{Baldo2000,Burin2000,Mashford2013}. The presence of molecular disorder is generally thought to hinder exciton transport properties~\cite{Bakalis1999,Markov2005,Lunt2010, Moix2013}, for instance by reducing diffusivity~\cite{Markov2005,Moix2013}; however, it has been speculated that in certain cases disorder might serve to enhance exciton transport~\cite{Mohseni2008,Moix2013}.

The first spectroscopic evidence of quantum coherence in photo-excited light harvesting complexes~\cite{Engel2007} inspired a flurry of studies aimed at understanding exciton transport in molecular systems~\cite{Collini2009,Nelson2011,Valleau2012}. Theoretical model studies using quantum master equation~\cite{Mohseni2008,Moix2013} and path-intergral formulation~\cite{Ray1999,Huo2010} have indicated that the highly efficient transport of excitation energy in biological and nanoscale systems can be attributed to an effective interplay between the two regimes of exciton transport: the local wave-like regime and the non-local hopping regime. When electronic coupling between molecular subunits is strong, excitons tend to delocalize and their dynamics involve the collective interplay of nuclear and electronic degrees of freedom. This type of exciton dynamics is referred to as \textit{coherent}, which differs qualitatively from the \textit{incoherent} dynamics that arise when intermolecular coupling is weak in which localized excitons undergo discrete hops between different molecular subunits~\cite{Grover1971,Yarkony1977,Scholes2003}. There are many examples of systems including some organic semiconductors,\cite{Turro1991,Pope1999,Bardeen2013} colloidal quantum dots,\cite{Alivisatos1996,Talapin2010,Nozik2008} and certain molecular aggregates~\cite{Fidder1991,Vlaming2013} where exciton transport is dominated by incoherent-type dynamics.

In photosynthetic systems, synergy between local excitonic coherence and thermal fluctuation from the environment results in energy transfer efficiency exceeding purely coherent or purely incoherent transfer.~\cite{Ringsmuth2012} The relative manifestation of these two regimes is governed by static and dynamic disorder.~\cite{Valleau2012,Moix2013,Chenu2015}. The effects of disorder in coherent exciton dynamics has been well studied in the context of general theoretical models. While static disorder has been shown to arrest diffusion via the phenomenon known as Anderson localization,~\cite{Anderson1958,Barford2009} it has also been shown that this effect can be removed by the presence of dynamic disorder, for example, in the form of a system-bath coupling.~\cite{Mohseni2008,Moix2013} Relatively little attention has been given to what role, if any, disorder plays in enhancing exciton transport in the incoherent regime.

In this article, we explore the relationship between static disorder and the dynamics of excitons in the context of a simple but general model for exciton diffusion in the limit of small intermolecular coupling. We demonstrate that although site energetic disorder leads to a reduction in the average steady state diffusivity of excitons, microscopic heterogeneity gives rise to regions of the material for which exciton mobility is enhanced relative to that in the absence of disorder. We go on to show that the heterogeneity in microscopic transport properties is correlated with local excitation energy, and that this heterogeneity is responsible for the time-dependent diffusivity that has been observed by transient photoluminescene spectroscopy~\cite{Crooker2002,Akselrod2014,Bardeen2014} as well as by more recently developed time-resolved optical microscopy~\cite{Akselrod2014,Akselrod2014B}. In the next section we describe our model for disordered incoherent exciton dynamics.

\section[sec0]{Model}
We employ a simple and general model of a semiconductor film composed of a collection of disordered molecular subunits. F\"orster theory provides the theoretical basis for our treatment of incoherent exciton dynamics~\cite{Forster1948}. In our model, excitons diffuse by performing a series of discrete hops between intermolecular subunits. Hopping rates are governed by the F\"orster rate equation,
\begin{equation}
k_{\rm{DA}}(d,\varepsilon_\mathrm{D},\varepsilon_\mathrm{A}) = \frac{1}{\tau }{\left( {\frac{{{R_0(\varepsilon_\mathrm{D},\varepsilon_\mathrm{A})}}}{d}} \right)^6},
\label{eqn:rate}
\end{equation}
where $d$ is the hopping distance, $\varepsilon_\mathrm{D}$ and $\varepsilon_\mathrm{A}$ are the absorption (excitation) energies of the donor and acceptor sites, respectively, $\tau$ is the experimentally determined exciton lifetime, and $R_0$ is the F\"orster radius. The dependence of exciton hopping rate on the excitation energies of molecular subunits is contained within the expression of the F\"orster radius,
\begin{equation}
{{R_0}(\varepsilon_\mathrm{D},\varepsilon_\mathrm{A})} = \left[\frac{9}{{8\pi }}\frac{{{c^4}{\hbar ^4}}}{{{n^4}}}\eta {\kappa ^2}\int {\frac{{{\sigma }\left( \varepsilon ; \varepsilon_A \right){f}\left( \varepsilon ; \varepsilon_D \right)}}{{{\varepsilon ^4}}}} d\varepsilon\right]^{1/6},
\label{eqn:radius}
\end{equation}
where $n$ is the refractive index, $c$ is the speed of light, $\hbar$ is the reduced Planck constant,  $\eta$ is the photoluminescence quantum yield, and $\kappa$ is the transition dipole orientation factor. The term in the integral contains the overlap between the normalized emission spectrum, $f(\varepsilon;\varepsilon_\mathrm{D})$, of the donor molecule with site energy, $\varepsilon_\mathrm{D}$, and the absorption spectrum, $\sigma(\varepsilon;\varepsilon_\mathrm{A})$, of the acceptor molecule with site energy $\varepsilon_\mathrm{A}$. The line shapes of $\sigma(\varepsilon;\varepsilon_\mathrm{A})$ and $f(\varepsilon;\varepsilon_\mathrm{D})$ are taken to be Gaussian with a mean of $\varepsilon_\mathrm{A}$ and $\varepsilon_\mathrm{D} - \Delta_\mathrm{ss}$, respectively, and each with a standard deviation equal to the homogeneous line width $\sigma_{\rm{h}}$. For a given molecular subunit, the difference in energy between the absorption and emission peaks is given by the Stokes shift, $\Delta_\mathrm{ss}$, which reflects the rapid (on the timescales of intermolecular excitonic transitions) electronic and nuclear relaxation that follows the excitation. Simulations are carried out in the dilute limit of exciton density.

We incorporate disorder into our model by varying the site energy (\textit{i.e.}, $\varepsilon_\mathrm{A}$ and $\varepsilon_\mathrm{D}$ in Eqs.~\ref{eqn:rate} and~\ref{eqn:radius}) of individual molecular subunits. For this study the model system consists of a two-dimensional array of 2500 hexagonally closed packed molecular subunits with absorption peak energies drawn randomly from a Gaussian distribution with mean $\bar{\varepsilon}$ and standard deviation $\sigma_\mathrm{ih}$, corresponding to the inhomogeneous linewidth. Each molecular subunit is assigned a fixed transition dipole vector (used to generate $\kappa$ in Eq.~\ref{eqn:radius}) oriented randomly on the surface of a unit sphere. The results presented here utilize parameters and a system geometry that were selected to correspond to the thin film of CdSe/ZnCdS core-shell quantum dots (the subject of recent experiments~\cite{Akselrod2014}). Nonetheless we expect the qualitative features of our model to apply generally across systems exhibiting incoherent exciton transport. The model system is described in greater detail in Ref. \citenum{Akselrod2014}. 

\section[sec1]{Local Exciton Hopping Rate Depends Primarily on Donor Site Energy}
To understand exciton transport from a microscopic perspective, we consider the phenomenological rate for exciton transfer from an individual electronically excited molecular subunit. For condensed phase systems which exhibit incoherent exciton transport, this rate is dominated almost entirely by the sum of nearest neighbor molecule-to-molecule transitions, which are uniform and constant in a perfectly ordered film but vary upon the introduction of disorder. In our model such variations arise from the dependence of Eqs.~\ref{eqn:rate} and \ref{eqn:radius} on energetic disorder, through the variables $\varepsilon_\mathrm{D}$, $\varepsilon_\mathrm{A}$, and spatial disorder, through the variables $d$ and $\kappa$. From the perspective of an electronically excited donor molecule, the local environment provides a sampling of the system heterogeneity. For multidimensional systems, the dependence of exciton transfer rate on the environmental variables $\varepsilon_\mathrm{A}$, $d$, and $\kappa$ are mitigated by the population of many nearest neighbors. The qualitative effect of site energetic disorder on exciton transport can therefore be understood primarily in terms of the donor site energy by considering the average molecule-to-molecule transfer rate from a molecular subunit with site energy $\varepsilon_\mathrm{D}$ to a nearest neighbor site, given by
\begin{equation}
\bar{k}(\varepsilon_\mathrm{D}) = \int_{-\infty}^\infty d\varepsilon_\mathrm{A} k_\mathrm{DA}(d_0,\varepsilon_\mathrm{D},\varepsilon_\mathrm{A}) P(\varepsilon_\mathrm{A}) %\approx k_\mathrm{DA}(d_0,\varepsilon_D,\bar{\varepsilon}),
\label{eqn:kbar}
\end{equation}
where $d_0$ is the nearest neighbor distance and $P(\varepsilon_\mathrm{A})$ is the distribution of site energies in the system,
\begin{equation}
P(\varepsilon_\mathrm{A})=\frac{1}{\sqrt{2 \pi {\sigma_\mathrm{ih}}^2}}\exp{\left(-\frac{\left(\varepsilon_\mathrm{A}-\bar \varepsilon\right)^2}{2{\sigma_\mathrm{ih}}^2 } \right)}
\label{eqn:P}
\end{equation}
which is in this case Gaussian with mean $\bar{\varepsilon}$ and variance $\sigma_\mathrm{ih}$. It turns out that even in two dimensions the physics contained within Eq.~\ref{eqn:kbar} are sufficient to understand the general qualitative effects of energetic disorder on exciton transport.

\begin{figure*}
\includegraphics{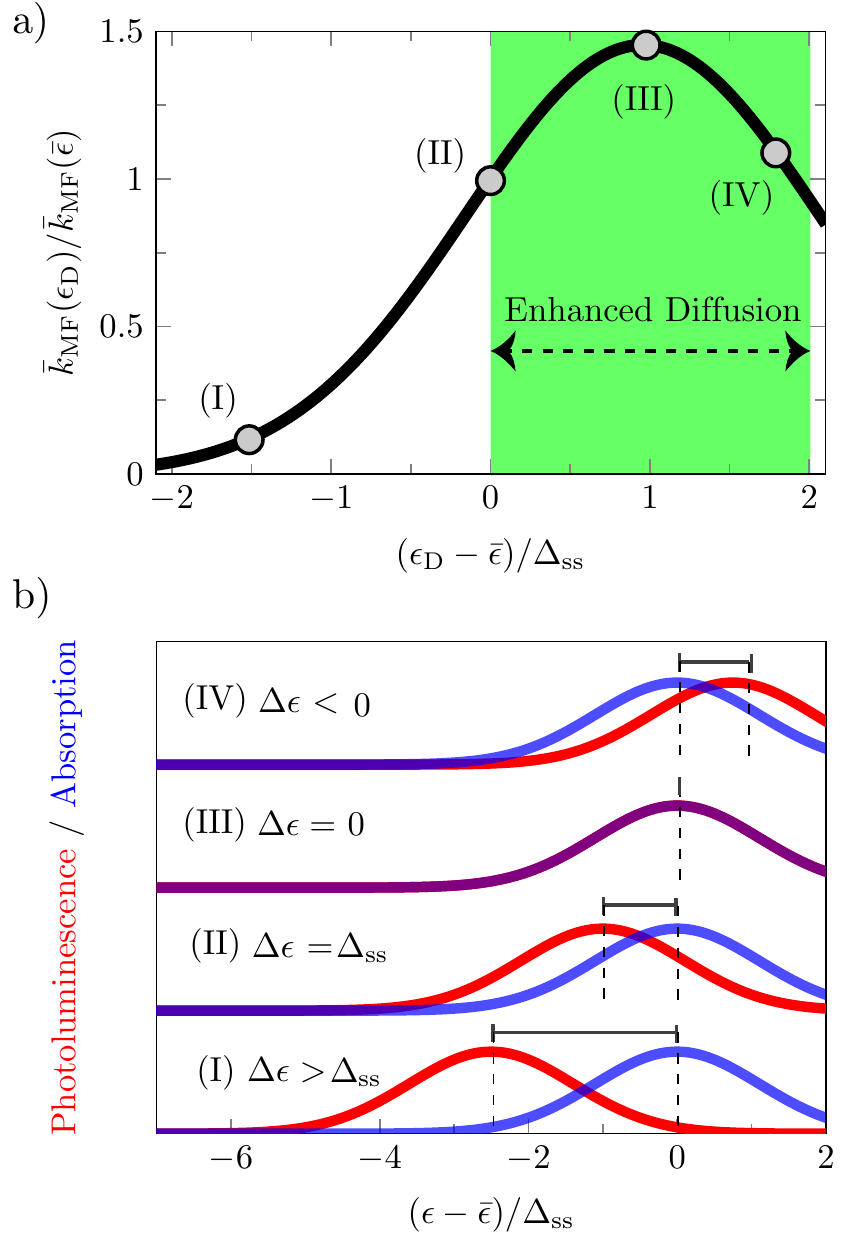}
\caption{(a) The average pairwise energy transfer rates, $\bar k_\mathrm{MF}(\varepsilon_\mathrm{D})$ between a donor molecule with site energy $\varepsilon_\mathrm{D}$ and an acceptor molecule with average site energy $\bar{\varepsilon}$. (b) The spectral overlap of the donor (red) and the acceptor (blue) subunits for selected values, numbered (I) through (IV), of $\bar k_\mathrm{MF}(\varepsilon_\mathrm{D})$ as shown in (a). The difference between the acceptor site energy (same as its absorption energy) and the donor emission energy (site energy offset by the Stokes shift, \textit{i.e.}, $\varepsilon_\mathrm{D} -\Delta_\mathrm{{ss}}$) is given by $\Delta \varepsilon = \bar \varepsilon - \varepsilon_\mathrm{D} +\Delta_\mathrm{{ss}}$.
Subunits with site energies between  $\bar{\varepsilon}$ and $\bar{\varepsilon} + 2 \Delta_\mathrm{ss}$, indicated by the shaded green region, exhibit exciton transfer rates that are enhanced relative to a perfectly uniform film. Hopping rates are calculated using Eq.~\ref{eqn:kbar} with $\sigma_{\rm{ih}} = 0.8\sigma_{\rm{h}}$, $\sigma_{\rm{h}}=42$ meV, $\Delta_{\mathrm{ss}} = 38$ meV, $n=1.7$, $\eta=1$, $d_0=8$ nm, $\kappa^{2}=\langle \kappa^{2} \rangle =2/3$, and $\tau=$ 10 ns.}
\label{fig:fig1}
\end{figure*}

For systems with symmetric and singly peaked energetic polydispersity, as is the case in this study, the expression in Eq.~\ref{eqn:kbar} can be simplified in the context of a mean field approximation to take the form $\bar{k}_\mathrm{MF}(\varepsilon_\mathrm{D})= k_\mathrm{DA}(d_0,\varepsilon_\mathrm{D},\bar{\varepsilon})$. Figure~\ref{fig:fig1}a contains a plot of $\bar{k}_\mathrm{MF}(\varepsilon_\mathrm{D})$, where it is shown that the mean exciton hopping rate is a non-monotonic function of donor site energy. The maximum transition rate is achieved when the donor site energy is above the average, specifically when $\varepsilon_\mathrm{D}  =  \bar{\varepsilon}+\Delta_{\mathrm{ss}}$, % beginning of bibnote
\footnote{
% \unexpanded{\makeatletter\let\@bibitemShut\relax\makeatother}
{For individual chromophores that have Gaussian homogeneous lineshapes, $k_\mathrm{DA}(d_0,\varepsilon_\mathrm{D},\bar{\varepsilon})$ can be derived as~\cite{Ahn2007,Akselrod2014}
%{equation*}
\[
k_\mathrm{DA}(d_0,\varepsilon_\mathrm{D},\bar{\varepsilon}) = \frac{C}{\sqrt{2}}{\left(\frac{2}{\varepsilon_\mathrm{D} + \bar \varepsilon - \Delta_\mathrm{ss}}\right)}^4\exp{\left({-\frac{{\left(\varepsilon_\mathrm{D} - \bar \varepsilon - \Delta_{\mathrm{ss}}\right)}^2}{4{\sigma_\mathrm{h}}^2 }}\right)},
\]
%\end{equation*}
where $C$ is a constant as defined in Eq.~\ref{eqn:C}. The maximum $k_\mathrm{DA}(d_0,\varepsilon_\mathrm{D},\bar{\varepsilon})$ occurs at $\varepsilon_{\mathrm{D}}=\Delta_{\mathrm{ss}}+\sqrt{{\bar{\varepsilon}}^2-8{\sigma_{\mathrm{h}}}^2}$. When $\sigma_\mathrm{h} \ll \bar{\varepsilon}$, the result simplifies to yield the optimal donor energy to be $\varepsilon_D \approx \Delta_{\mathrm{ss}} + \bar{\varepsilon}$.}} 
% end of bibnote
where the magnitude of the donor-acceptor spectral overlap integral in Eq.~\ref{eqn:radius} is optimized. (See Fig.~\ref{fig:fig1}b.) 

The average hopping rate of a material reflects the distribution of local hopping rates. In the next section we evaluate the system-wide average hopping rate and discuss how it depends on the details of static disorder.

\section{The Effect of Disorder on the Material-Wide Average Exciton Hopping Rate}
The net effect of site energetic disorder on the material-wide exciton hopping rates can be evaluated by comparing films with and without energetic disorder. From Eqs.~\ref{eqn:radius} and~\ref{eqn:kbar}, we see that in the absence of energetic disorder (\textit{i.e.}, $P(\varepsilon_\mathrm{A}) = \delta(\varepsilon_\mathrm{A} - \bar{\varepsilon})$), the nearest-neighbor exciton transfer rate is constant and is equal to
\begin{eqnarray}
k_\mathrm{DA}\left(d_0,\bar \varepsilon, \bar \varepsilon \right)&=&\frac{1}{\tau}\left(\frac{{R_0(\bar \varepsilon,\bar \varepsilon)}}{{d_0}}\right)^6 
\nonumber \\
&=& C\int_{-\infty}^\infty d\varepsilon \frac{1}{\varepsilon^4}\exp{\left(-\frac{\left(\bar \varepsilon-\varepsilon\right)^2}{2{\sigma_\mathrm{h}}^2 } \right)}\frac{1}{\sqrt{2 \pi {\sigma_\mathrm{h}}^2 }}\exp{\left(-\frac{\left(\bar \varepsilon-\Delta_\mathrm{ss} - \varepsilon\right)^2}{2{\sigma_\mathrm{h}}^2 } \right)}
\label{eqn:2} \\
&=&\frac{C}{\sqrt{2}}\frac{1}{\left(\bar \varepsilon - \frac{\Delta_\mathrm{ss}}{2}\right)^4}\exp{\left({-\frac{\Delta_{\mathrm{ss}}^2}{4{\sigma_\mathrm{h}}^2 }}\right)},
\label{eqn:k_avg}
\end{eqnarray}
where $C$ is a collection of physical constants, 
\begin{equation}
C \equiv \frac{9}{{8\pi }}\frac{{{c^4}{\hbar ^4}}}{{{n^4}}}{\eta}{\sigma (\bar \varepsilon)}\langle  {\kappa ^2}\rangle\frac{1}{\tau {d_0}^6},
\label{eqn:C}
\end{equation}
which includes the absorption cross section $\sigma$ at $\bar \varepsilon$, $\sigma (\bar \varepsilon)$, and the average dipole orientation factor, $\langle  {\kappa ^2}\rangle = 2/3$. The expression in Eq.~\ref{eqn:k_avg} reflects the the assumption that the ${\varepsilon^4}$ term does not vary significantly over the Gaussian term with respect to $\varepsilon$ in Eq.~\ref{eqn:2},~\cite{Ahn2007} which is valid in cases where the linewidth is smaller than the band gap.

For a film \textit{with} energetic disorder, the nearest-neighbor exciton transfer rate depends on exciton energy, as illustrated in Fig.~\ref{fig:fig1}a. Notably, the molecular subunits with site energies between  $\bar{\varepsilon}$ and $\bar{\varepsilon} + 2 \Delta_\mathrm{ss}$ exhibit exciton transfer rates that are enhanced relative to a perfectly ordered film. However, a consequence of the non-monotonic turnover in exciton transfer rates for $\varepsilon_\mathrm{D}>\bar{\varepsilon}+\Delta_\mathrm{ss}$ is that the system-wide average exciton transfer rate is expected to be reduced relative to that of an energetically ordered system. More formally, the film-wide average exciton transfer rate is given by
\begin{eqnarray}
%\langle \bar{k}(\varepsilon_\mathrm{D})\rangle = \int_{-\infty}^\infty \int_{-\infty}^\infty d\varepsilon_\mathrm{D} d\varepsilon_\mathrm{A} k_\mathrm{DA}(d_0,\varepsilon_\mathrm{D},\varepsilon_\mathrm{A}) P(\varepsilon_\mathrm{A}) P(\varepsilon_\mathrm{D}), 
\langle k \rangle &=& \int_{-\infty}^\infty d\varepsilon_\mathrm{D} \bar{k}(\varepsilon_\mathrm{D}) P(\varepsilon_\mathrm{D}), \nonumber \\
&=&\int_{-\infty}^\infty \int_{-\infty}^\infty d\varepsilon_\mathrm{D} d\varepsilon_\mathrm{A} k_\mathrm{DA}(d_0,\varepsilon_\mathrm{D},\varepsilon_\mathrm{A}) P(\varepsilon_\mathrm{A}) P(\varepsilon_\mathrm{D}). 
\label{eqn:kbar_eq}
\end{eqnarray}
Since $P(\varepsilon_D)$ and $P(\varepsilon_A)$ are Gaussian density of states centered at $\bar \varepsilon$ for both donors and acceptors as defined in Eq.~\ref{eqn:P}, Eq.~\ref{eqn:kbar_eq} can be rewritten as
\begin{eqnarray}
\langle {k}\rangle= \frac{C}{\left( 2 \pi \right )^{3/2}{\sigma_\mathrm{ih}}^2{\sigma_\mathrm{h}} } \int_{-\infty}^\infty \int_{-\infty}^\infty \int_{-\infty}^\infty \frac{d\varepsilon_\mathrm{A} d\varepsilon_\mathrm{D} d\varepsilon}{\varepsilon^4}  \exp{\left(-\frac{\left(\varepsilon_\mathrm{A}-\bar \varepsilon\right)^2}{2{\sigma_\mathrm{ih}}^2 } \right)} \nonumber \\
\times \exp{\left(-\frac{\left(\varepsilon_\mathrm{A}-\varepsilon\right)^2}{2{\sigma_\mathrm{h}}^2 } \right)} \exp{\left(-\frac{\left(\varepsilon_\mathrm{D}-\bar \varepsilon\right)^2}{2{\sigma_\mathrm{ih}}^2 } \right)}\exp{\left(-\frac{\left(\varepsilon_\mathrm{D}-\Delta_\mathrm{ss} - \varepsilon\right)^2}{2{\sigma_\mathrm{h}}^2 } \right)}.
\label{eqn:final1}
\end{eqnarray}
To evaluate Eq.~\ref{eqn:final1}, we first integrate with respect to $\varepsilon_\mathrm{A}$ and $\varepsilon_\mathrm{D}$ and then $\varepsilon$ because the terms involving $\varepsilon_\mathrm{A}$  and $\varepsilon_\mathrm{D}$ are independent from one another while $\epsilon$ is dependent on other two variables. This leads to 
\begin{equation}
\langle {k}\rangle= \frac{C}{\sqrt{ 2 \pi {\sigma_\mathrm{h}}^2 }} \frac{ 1 } {\beta^2+1} \int_{-\infty}^\infty \frac{d\varepsilon}{\varepsilon^4} \exp{\left[-\frac{\left(\varepsilon - \left(\bar \varepsilon - \frac{\Delta_\mathrm{ss}}{2}  \right) \right)^2}{\left(\beta^2 + 1\right){\sigma_\mathrm{h}}^2} \right]} \exp{\left[-\frac{{\Delta_\mathrm{ss}}^2}{4\left(\beta^2 + 1\right){\sigma_\mathrm{h}}^2} \right]},
\label{eqn:final2}
\end{equation}
where $\beta \equiv \sigma_\mathrm{ih} / \sigma_\mathrm{h}$ is the ratio of static disorder to the dynamic disorder. Again assuming that the ${\varepsilon^4}$ term does not vary significantly over the Gaussian term with respect to $\varepsilon$ in Eq.~\ref{eqn:final2}, we can simplify the above equation as
\begin{equation}
\langle {k}\rangle= \frac{C}{\sqrt{ 2}} \frac{1}{\sqrt{\beta^2+1}}\frac{1}{\left(\bar \varepsilon - \frac{\Delta_\mathrm{ss}}{2}\right)^4}\exp{\left[-\frac{{\Delta_\mathrm{ss}}^2}{4\left(\beta^2 + 1\right){\sigma_\mathrm{h}}^2} \right]},
\label{eqn:kbar_avg_beta}
\end{equation}
which can be rewritten in the terms of the original variables as
\begin{equation}
\langle {k}\rangle=  \frac{C}{\sqrt{ 2}} \frac{{\sigma_\mathrm{h}}}{\sqrt{{\sigma_\mathrm{ih}}^2+{\sigma_\mathrm{h}}^2}}\frac{1}{\left(\bar \varepsilon - \frac{\Delta_\mathrm{ss}}{2}\right)^4}\exp{\left[-\frac{{\Delta_\mathrm{ss}}^2}{4\left({\sigma_\mathrm{ih}}^2+{\sigma_\mathrm{h}}^2\right)} \right]}.
\label{eqn:kbar_avg}
\end{equation}
By comparing Eqs.~\ref{eqn:k_avg} and~\ref{eqn:kbar_avg}, we find that the average exciton transfer rate is decreased by disorder (\textit{i.e.}, $\langle k \rangle < k_\mathrm{DA}\left(d_0,\bar \varepsilon, \bar \varepsilon \right)$) if
\begin{equation}
\left(\frac{{\Delta_\mathrm{ss}}}{{\sigma_\mathrm{h}}}\right)^2 < \frac{2(1+\beta^2)}{\beta^2}\ln{(1+\beta^2)}.
\label{eqn:condition1}
\end{equation}
The right hand side of Eq.~\ref{eqn:condition1} is a monotonically increasing function of $\beta^2$, and also is greater than or equal to 2, with the equality occurring when $\beta=0$. Therefore, for \textit{any} non-zero value of $\beta$, the average exciton hopping rate for an energetically disordered film can only be enhanced relative to that of an ordered film (\textit{e.g.}, $\langle k \rangle > k_\mathrm{DA}\left(d_0,\bar \varepsilon, \bar \varepsilon \right)$) when
\begin{equation}
\left(\frac{{\Delta_\mathrm{ss}}}{{\sigma_\mathrm{h}}}\right)^2 \ge 2.
\label{eqn:condition2}
\end{equation}
The value on the right hand side of Eq.~\ref{eqn:condition1} grows with $\beta$ and is thus larger than 2 for typical systems (\textit{e.g.}, 2.5 for the system considered here in which $\beta = 0.8$). Equations~\ref{eqn:condition1} and~\ref{eqn:condition2} tell us that energetic disorder only serves to enhance the film-wide average exciton transfer rate when the ordered system has poor donor-acceptor spectral overlap, in which case disorder serves to facilitate energetic resonance between some nearest neighbor pairs. An analogous type of analysis can be carried out with respect to spatial disorder in which the molecule-to-molecule transition rates are modulated through the $d$-dependence (or $\kappa$-dependence) of Eq.~\ref{eqn:rate}. The results are analogous to those described above; the presence of spatial disorder typically leads to a reduction in the film-wide average exciton transfer rate.

The observed rate of exciton transfer is only equal to the material-wide average when excitons are are uniformly distributed across the ensemble of molecular subunits. Energetic disorder, however, has another important effect on exciton transport in that it facilitates the time-varying dissipation of excitation energy, thereby biasing excitons towards lower-energy molecular subunits, those with reduced effective hopping rates. (See Fig.~\ref{fig:fig1}a.) In the next section we employ Monte Carlo simulations to explore time-dependent effect of energetic relaxation on observed exciton diffusivity.

\section{Transient Energetic Relaxation Leads to Time Dependent Exciton Diffusivity}

Each energy transfer event is associated with small energetic losses. In our model, the dissipative mechanism appears in the form of the Stokes shift, $\Delta_\mathrm{ss}$, designed to mimic the rapid internal electronic and nuclear relaxation that follows molecular excitation. Because of the Stokes shift, the most probable transitions---those which maximize overlap between donor emission and acceptor absorption line shapes---are to acceptor molecules with excitation energies that are $\Delta_\mathrm{ss}$ lower than that of the donor molecule. Therefore, the initial diffusion of randomly populated excitons is accompanied by a decrease in average exciton energy. 

\begin{figure}
  \includegraphics{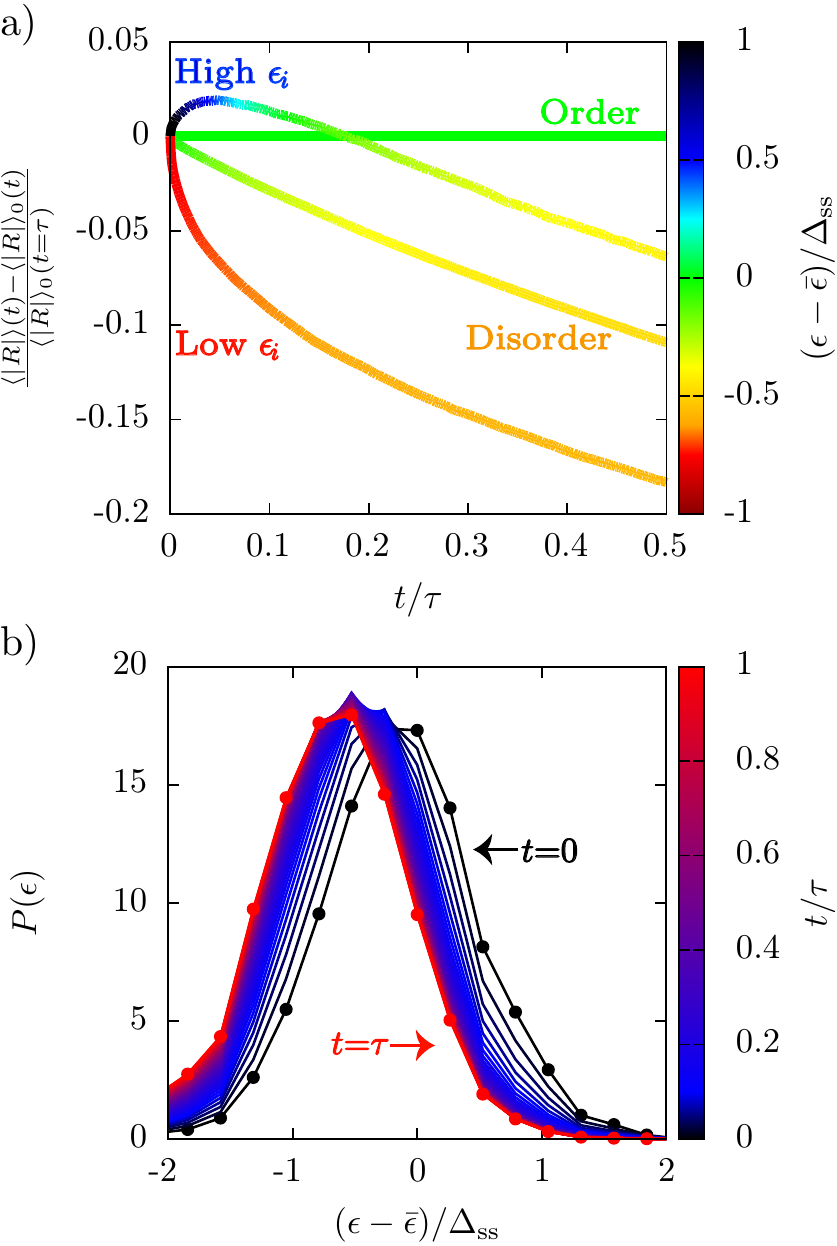}
  \caption{Exciton dynamics within energetically disordered molecular film. (a) The average diffusion length of excitons in a disordered film plotted relative to those in an energetically ordered film, the latter denote by ${\left\langle {\left| R \right|} \right\rangle_0}(t)$. The four curves correspond to the ordered film result, an average over excitons initialized uniformly across within a disordered film (the curve labeled \textit{disorder}) and an average over the subpopulation of excitons initialized on either high energy (labeled high $\varepsilon_i$) or low energy (labeled low $\varepsilon_i$) sites. This quantity has been normalized by the diffusion length in the ordered film measured at the exciton's lifetime (${\left\langle {\left| R \right|} \right\rangle_0} (t = \tau)$). The average site energy visited by the exciton is indicated by its correpsonding color. (Blue is high and red is low energy.) (b) Time-evolution of the probability density distribution of excitons with energy $\varepsilon$. Time points are indicated by its corresponding color. (Black is the initial state, and red is the state in which most excitons have already decayed.)}
  \label{fig:fig2}
\end{figure}

Using a standard kinetic Monte Carlo algorithm we have studied the dynamics of excitons by analyzing the trajectories of individual non-interacting excitons, with a particular focus on the relationship between exciton energy and diffusivity. Figure~\ref{fig:fig2}a compares the time dependent average displacement of a uniform distribution of excitons in an energetically disordered film to that in a film without energetic disorder. We observe that excitons within the disordered film (see the curve labeled as `disorder' in Fig.~\ref{fig:fig2}a) exhibit reduced diffusivity when compared to a perfectly ordered film (see the curve labeled `uniform' in Fig.~\ref{fig:fig2}a). In addition, we show the average displacement for the subpopulation of excitons initialized on high energy molecular subunits, those with excitation energies $\varepsilon_i= \bar \varepsilon + \Delta_\mathrm{ss}$, or low energy molecular subunits, those with excitation energies $\varepsilon_i = \bar \varepsilon - 0.8\Delta_\mathrm{ss}$. We observe that the initial behavior of either subpopulation is consistent with the results plotted in Fig.~\ref{fig:fig1}a, namely that excitons with low initial energies diffuse more slowly than the ensemble average while excitons with high initial energies diffuse more rapidly than the ensemble average. The high energy excitons even manifest a short time enhancement in diffusivity relative to that of an energetically uniform film. The enhancement in diffusion for high-energy excitons is short lived because this population of excitons rapidly thermalizes, preferring to occupy lower energy molecular subunits whose effective hopping rates no longer reflect their initial enhancement. (See Fig.~\ref{fig:fig1}a.) Consequently, over timescales that permit multiple site-to-site transitions, the average distance traveled by excitons, regardless of their initial energy, becomes less than the average distance traveled by excitons in the absence of energetic disorder. 

Evidence for the rapid dissipation of energy for the \textit{diffusion-enhanced} excitons can be seen by analyzing the transient energetics of simulated exciton trajectories (represented by line color in Fig.~\ref{fig:fig2}a). Specifically, the line color in Fig.~\ref{fig:fig2} corresponds to the mean site energy, averaged over a given population of excitons, at a specific time $t$. In contrast to the rapid energetic decrease observed in the population of excitons with initially high energies, excitons with initially low energies exhibit an initial increase in energy, indicating that the lowest energy sites are not traps but are dynamically repopulated. The transient energetics can be understood more deeply by considering the time-dependent distribution of exciton energies (see Fig.~\ref{fig:fig2}b). 

In Figure~\ref{fig:fig2}b we plot $P(\varepsilon)$, the probability that an exciton resides on a molecular subunit with excitation energy $\epsilon$, resolved as a function of time, where at $t=0$ excitons are uniformly distributed across the all molecular subunits. We observe a pronounced transient red-shift starting from an initially uniform distribution (black line in Fig.~\ref{fig:fig2}b) to a final near steady state distribution (red lines in Fig.~\ref{fig:fig2}b), the result of individual excitons performing energy lowering molecule-to-molecule transitions. During the non-equilibrium relaxation, the population of excitons residing on molecular subunits with enhanced hopping rates (relative to systems without energetic disorder) decrease rapidly with time, which is accompanied by a corresponding increase in the population of excitons that reside on sites with reduced hopping rates (relative to a system without energetic disorder). The combined effect is a time-dependent lowering of net exciton diffusivity as more excitons occupy molecular subunits with low energy and  reduced inter-molecular transition rates.

\begin{figure}
  \includegraphics{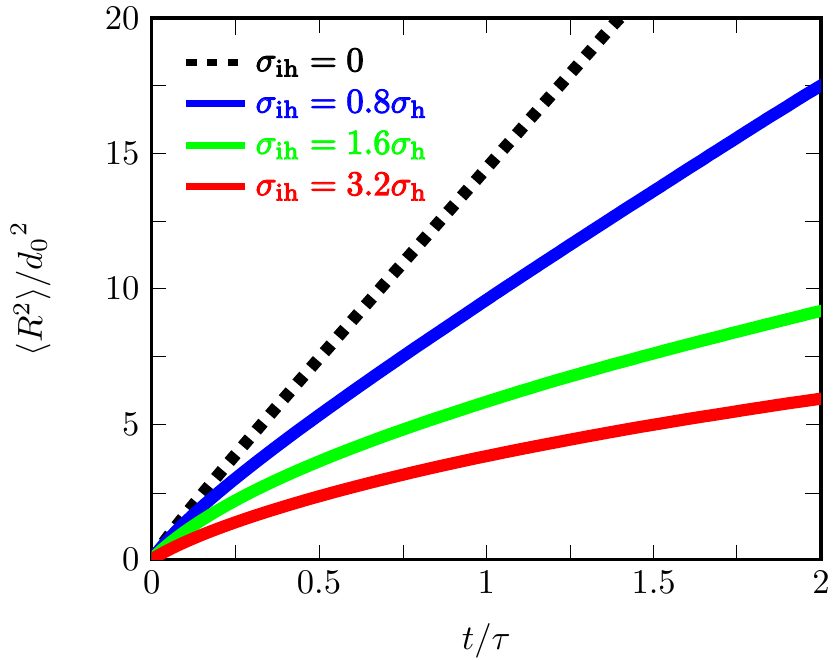}
  \caption{Mean square displacements (MSD), $\langle R^2 \rangle$, of excitons in molecular films with various degrees of site energetic disorder. MSD is normalized by the square of the nearest neighbor distance, $d_0$, such that $\langle R^2 \rangle/{d_0}^2$ is proportional to the square of the number of exciton hops. In the presence of static disorder, $\langle R^2 \rangle$ is shorter than that of the uniform energy case, $\sigma_\mathrm{ih}=0$, at all times.}
  \label{fig:fig3}
\end{figure}

The relationship between static energetic disorder and transient exciton energetics has been previously explored in detail by B\"assler~\cite{Bassler1993}. The phenomenon is experimentally measurable through the time-dependence of the fluorescence spectrum~\cite{Crooker2002,Akselrod2014}. Similarly, the relationship between static energy disorder and exciton transport is experimentally accessible through the time- and position-dependence of the fluorescence spectrum~\cite{Madigan2006,Akselrod2014,Akselrod2014B} where the transient lowering of net exciton diffusivity manifests itself in the mean squared displacement (MSD), $\langle R^2 \rangle$. In the absence of disorder (\textit{i.e.}, $\sigma_\mathrm{ih}=0$), exciton transport is diffusive (\textit{i.e.}, $\langle R^2 \rangle \propto t$) at all times. % beginning bibnote
\footnote{{As shown in Ref.~\citenum{Ahn2007}, MSD grows non-linearly initially in the presence of randomly oriented transition dipoles, giving rise to a subdiffusive transport even in the absence of static disorder. However, this non-linear effect appears to be small in the time-scale plotted in Fig.~\ref{fig:fig3} such that the MSD grows linearly with time ($\langle R^2 \rangle \propto t^\alpha$ where $\alpha \approx 1$).}}
%%% end bibnote
On the other hand, static disorder gives rise to subdiffusive transport~\cite{Fa2003,Ahn2007,Akselrod2014} (\textit{i.e.}, $\langle R^2 \rangle \propto t^\alpha$ where $\alpha < 1$) with a MSD that grows non-linearly at short times and only becomes linear at longer times when the system has reached a dynamic equilibrium. As shown in Fig.~\ref{fig:fig3}, the net effect of disorder is to reduce the mean squared displacement of excitons, especially at long times. Additionally, the extent of initial deviations from the $\sigma_\mathrm{ih}=0$ case depends sensitively on the extent of static disorder, as does the steady state diffusion constant (\textit{i.e.}, the slope of $\langle R^2 \rangle$ at long times).
%Additionally, the extent of initial deviations from the $\sigma_\mathrm{ih}=0$ case and the steady state diffusion constant (\textit{i.e.}, the slope of $\langle R^2 \rangle$ at long times) depend sensitively on the extent of static disorder.

\section{Conclusions}
In summary, using a simple F\"orster-type model of incoherent exciton dynamics in an energetically disordered array of molecular subunits, we have shown that site-energetic disorder leads to variations in transport properties of individual molecular subunits. The variations correlate with site energy, which can be understood in terms of spectral overlap integrals, and have the net effect of reducing the average molecule-to-molecule transition rate relative to that of a perfectly ordered material. Furthermore, in the presence of disorder, the steady state distribution of exciton energies is not identical to the distribution of site energies due to the dissipation of electronic energy that accompanies typical molecule-to-molecule transitions. Consequently, the steady state distribution of excitons is biased towards sites with lower than average excitation energies, and thus with exciton transfer rates that are reduced relative to that in an ordered material. Hence the observed exciton diffusivity is further lowered by transient exciton diffusion. The answer to the the question posed in the title of this article is length and time scale dependent. For macroscopic systems the answer is \textit{no}: the average effect of disorder is to reduce overall exciton transport. However, at the nanoscale the answer is \textit{yes}: disorder can serve to enhance exciton mobility in certain microscopic regions. Such local enhancement may have played a role in optimizing the structure of chromophore network in light harvesting complexes.~\cite{Ringsmuth2012} Considering such microscopic enhancements (along with their diminishing counterparts) will become increasingly important as devices continue to incorporate  nanotechnological elements.

\section*{Acknowledgement}
E.M.Y.L thanks Chee Kong Lee and Liang Shi for helpful discussions. The authors acknowledge support from the National Science Foundation Graduate Research Fellowship Program. 
%\end{acknowledgement}

% supplementary info
%\begin{suppinfo}
%Derivations for Eqs.~\ref{eqn:kbar_avg} and~\ref{eqn:k_avg}.
%\end{suppinfo}

%\bibliographystyle{achemso}
%\bibliographystyle{plainnat}
%\bibliographystyle{aipauth4-1}
\bibliography{draft_ref}

\end{document}